# Title

Online geometric calibration of a hybrid CT system for ultrahigh-resolution imaging

# Authors

Dakota H. King, Eric E. Bennett, Dumitru Mazilu, Muyang Wang, Marcus Y. Chen, Han Wen

Laboratory of Imaging Physics, Biochemistry and Biophysics Branch, Division of Intramural Research, National Heart, Lung and Blood Institute, National Institutes of Health, USA;

Correspondence: han.wen@nih.gov; Tel.: +1 301-496-2694

# Abstract

A hybrid imaging system consisting of a standard CT scanner and a low-profile photon-counting detector insert in contact with the patient's body has been used to produce ultrahigh-resolution images in a limited volume in chest scans of patients. The detector insert is placed on the patient bed as needed and not attached. Thus, its position and orientation in the scanner is dependent on the patient's position and scan settings. To allow accurate image reconstruction, we devised a method of determining the relative geometry of the detector insert and the CT scanner for each scan using fiducial markers. This method uses an iterative registration algorithm to align the markers in the reconstructed volume from the detector insert to that of the concurrent CT scan. After obtaining precise geometric information of the detector insert relative to the CT scanner, the two complementary sets of images are summed together to create a detailed image with reduced artifacts.

# Introduction

### The hybrid CT scanner

The hybrid CT scanner consists of a standard computed tomography (CT) scanner and an additional low-profile photon-counting detector (the contact detector insert, or CDI) placed on the patient bed in contact with the patient's chest (Fig. 1). During a scan, the CT scanner's detector in the rotating gantry and the stationary CDI acquire data simultaneously. The hybrid system allows for ultrahigh-resolution imaging in a limited volume in front of the CDI, as the penumbra of the x-ray source on the CDI is minimized by its close proximity to the patient's body. The design, operation and testing of the hybrid system are described in detail elsewhere[1], [2]. The focal spot penumbra is usually the limiting factor of resolution in body scans in current clinical scanners. This is because high levels of x-ray power are required to achieve short scan times, resulting in finite focal spot size. Using this hybrid method, spatial resolution of 150 $\mu$m has been demonstrated in a standard CT scanner[1], [2].

In the rest of the paper, images reconstructed from the CDI-acquired data are referred to as "CDI images", and the concurrent images from the CT scanner itself are referred to as the "CT images". The CDI portion operates in the tomosynthesis mode[3], similar to a digital breast tomosynthesis scanner[4], [5].

### *The need for online geometric calibration*

Geometric calibration is motivated by two primary factors: tomosynthesis reconstruction of the CDI data, and the ability to accurately fuse the reconstructed volumes from the CDI and the CT scanner. Fusion of the images from the two sources require alignment of the two reconstructed volumes (Fig. 1C). Additionally, tomosynthesis reconstruction of the CDI data requires knowing the position of the x-ray focal spot relative to the CDI at every time point. Both aspects depend on knowing the geometry of the CDI in the CT scanner's bore. Errors in this information will result in misalignment artifacts in the CDI images, and disagreement in the spatial position of features in the CDI images versus the CT images. Because the position and orientation of the CDI vary with the patient's weight distribution and posture on the patient bed, they need to be determined for each scan, that is, online calibration is necessary.

### *What is new about the present geometric calibration method*

A number of calibration methods have been developed for cone-beam tomosynthesis as well as computed tomography (CT) and to meet the demand of diverse applications[6]–[41]. They have been generally classified into offline phantom-based calibration methods and online phantom-less methods[28]. Examples of recent literature reviews are provided by Graetz[39] and Jiang et al.[34]. In offline methods, the geometry of the imaging system is determined with dedicated calibration scans of phantoms. The phantoms contain distinct positional markers, such as radio-opaque beads[19], [22], [23], [30], [31], [35], [36], [41] or wires[33]. Some phantom-based methods do not need to know the precise arrangement of the markers by multiple scans of the same phantom in different positions[9], [11], [23], [24], [36], [39]. Online calibration methods are used for CT scans with projection angles spanning 180 to 360 degrees. They are phantom-less[7], [16]–[18], [20], [28], [29]. These methods can rely on quantifying misalignment using qualities of the reconstructed images itself: artifacts and image blurring that occur from error in geometric parameters [17], [20]. Other methods rely on exploiting the symmetry in a full 360-degree scan [7], [16], [28].

However, in contrast to existing online calibration methods that are designed for full-rotation CT scans, data acquired by the CDI have a truncated range of projection angles. The truncated range of projection angles causes image artifacts itself, which makes it difficult to isolate and quantify the level of artifacts that are associated with calibration errors. On the other hand, a helpful circumstance of the hybrid system is that half of the system, the commercial CT scanner, is fully calibrated, and the CDI detector itself is visible in the CT images. Therefore, we designed a novel method custom to the needs of this system. It is an online method with characteristics of a phantom-based method. Fiducial point markers are placed around the outside of the patient's body, in the z-range covered by the CDI. The true positions of the markers are determined for each scan from the CT images. This then allows us to construct and minimize a cost function that measures the difference between fiducial marker positions in the CDI and the CT images.

## Methods

### Imaging Procedure and the Anthropomorphic Phantom

The hybrid system consisted of a low-profile photon-counting detector as the CDI inside a commercial clinical CT scanner. The CDI has a 512 x 4100-pixel matrix (100 um pixel dimension) and can acquire data at a rate of 2,000 frames/s. The start of the CDI data acquisition was synchronized to the switch-on of the x-ray beam in the CT scanner via the external light indicator of the beam. The CT scan parameters were 120 kV/500 mA, axial (static bed) mode with a single rotation of 1.5 s period and a detector collimation of 120x0.5mm.

A Kyoto Kagaku PH-1 LUNGMAN chest phantom was imaged. The LUNGMAN phantom is anthropomorphic in that it simulates an adult human chest with detailed structures of synthetic soft tissue, lung vasculature and bone, each with x-ray absorption comparable to that of real tissue. Fiducial markers were seven tungsten carbide beads of 1 mm diameter. They were attached to a plastic strip in a linear arrangement, which was wrapped around the outer surface of the phantom (Fig. 2A). The precise arrangement of the beads was not critical, as long as they were all visible in the CDI images.

*Initial Estimation of the Geometric Parameters*

The geometric parameters to be determined are the position and orientation of the CDI in the CT scanner's world coordinate system, and the angle of the x-ray focal spot around its circular trajectory at each time point of the scan.

A method for finding an initial approximation of these parameters was to estimate the position of the CDI inside the CT scanner bore based on CT images that included the CDI. These estimates were imprecise, because the CT images had strong metal artifacts from the CDI's metal parts and its radiation shields (Fig. 2C). The second part of the initial approximation was the angle of the x-ray focal spot at each time point, based on determining the "sunrise" and "sunset" frames in the projection images acquired by the CDI. These were the time points where the x-ray focal spot aligned with the CDI's image plane, either on its rise above the image plane or fall below the image plane. They were determined visually and imprecise due to several factors, including scattered x-rays that blurred the transition of the image intensity at sunrise and sunset, the finite frame rate of the CDI and the finite focal spot size. In practice, the initial solution was sufficient for an initial tomosynthesis reconstruction of the CDI data, although the alignment of the reconstructed CDI images with the CT images was imprecise.

*Locating the Fiducial Markers in the CT Images*

Before finding positions of the fiducial markers in the CT images, 25.39 mm FOV sub-volumes (50 um pixels) with a 0.25 mm slice thickness were reconstructed around each bead using standard software on the CT scanner. Each sub-volume was analyzed to find the marker position. First, the region around the marker was segmented out by intensity thresholding at the level of half the maximum intensity (Fig. 2D). The segmentation only kept voxel values corresponding to the 3D intensity profile of the fiducial marker. Second, the center-of-mass of this segmented profile was computed to find the position of the fiducial marker. Third, the position coordinates were transformed to the scanner's world coordinate system using DICOM header information of the sub-volume. These fiducial marker positions were used as the "true" positions to align the CDI's reconstructed volume. To find the positions of markers in the CDI's reconstructed volume, the intensity thresholding and center-of-mass were computed in the same way as with the CT images.

*Iterative Calibration Algorithm*

The calibration concerns three coordinate systems: the CT scanner's world system, the scanner's image system, and the CDI's coordinate system. The relationship between the CT's world and image systems is known and available from the DICOM headers of the reconstructed images by the scanner. Their geometric relationship to the CDI's coordinate system needs to be determined for each scan. Additionally, the x-ray focal spot travels on a circular trajectory in the CT world system at a known radius and angular speed, but with an unknown angle offset at the start of each scan. Therefore, the calibration method is aimed at finding 7 parameters, $\vec{p}$ (3 rotational; 3 translational) of the 4x4 transformation matrix, $T_{CT,CDI}$, that maps a 3D point in the CDI system to that of the CT image system, and the angle offset of the x-ray focal spot at the start of the scan, $\varphi_0$.

Starting from the initial estimates $T_{CT,CDI}$ and $\varphi_0$ which are described in the previous section "Initial estimation of the geometric parameters", the trajectory of the x-ray focal spot was transformed into the CDI's coordinate system. The projection images from the CDI were then used to reconstruct sub-volumes containing the marker beads via weighted filtered back-projection (WFBP). Using the transformation matrix $T_{CT,CDI}$, these sub-volumes were defined in the CT image system and had the same exact voxel positions as the reconstructed sub-volumes from the CT scanner (see previous section "Locating the fiducial markers in the CT images"). This allows a cost function to be created that quantifies the amount of errors in the alignment of the bead positions from the two sources:

$$C(\vec{p}, \varphi_0) = \frac{1}{N} \sum_{i=1}^{N} |\vec{r}_{CT,i} - \vec{r}_{CDI,i}|^2.$$

Where $\vec{r}_{CT,i}$ and $\vec{r}_{CDI,i}$ are the positions of the marker beads in the CT scanner's reconstructed volume and the CDI's reconstructed volume, respectively. Each of the 7 parameters was then altered in turn, and the WFBP reconstruction of the CDI data was repeated to update the cost function. The process iterated multiple times through the 7 parameters until the cost function C was minimized via the Powell's method [42]. The pre- and post-iteration cost function values were compared to quantify improvement in the alignment of the beads.

To reduce the amount of time required for each CDI reconstruction iteration, the multiprocessing and multithreading packages in Python 3.8 were used. Projection images were split into groups where one python process was responsible for one group. Upon completion of one group, depending on how many groups were needed to compute, the process would continue to work on other groups. In addition to running multiple processes, each process made use of multiple threads to further reduce computation time.

*Quantifying Alignment Using Blood Vessels*

For additional quantification of alignment, 2D positions of small vessel structures from the CDI and CT images were compared at several image slices spanning the reconstructions. Contours were defined around each vessel using a threshold of half the maximum intensity of the vessel; the center of mass of the pixel values within the contour was used to calculate the position. Small vessels, 2 mm or smaller in diameter, were chosen for this analysis. Ten total vessels were analyzed: three located in slices in the +z direction, three in the -z direction, and four near the middle slice. In addition, the vessels chosen were distributed at different regions in the pixel grid (x- and y-coordinates) to represent various regions of the phantom. Pre- and post-calibration errors were compared to give an additional quantification and verification of alignment throughout the reconstructed volumes.

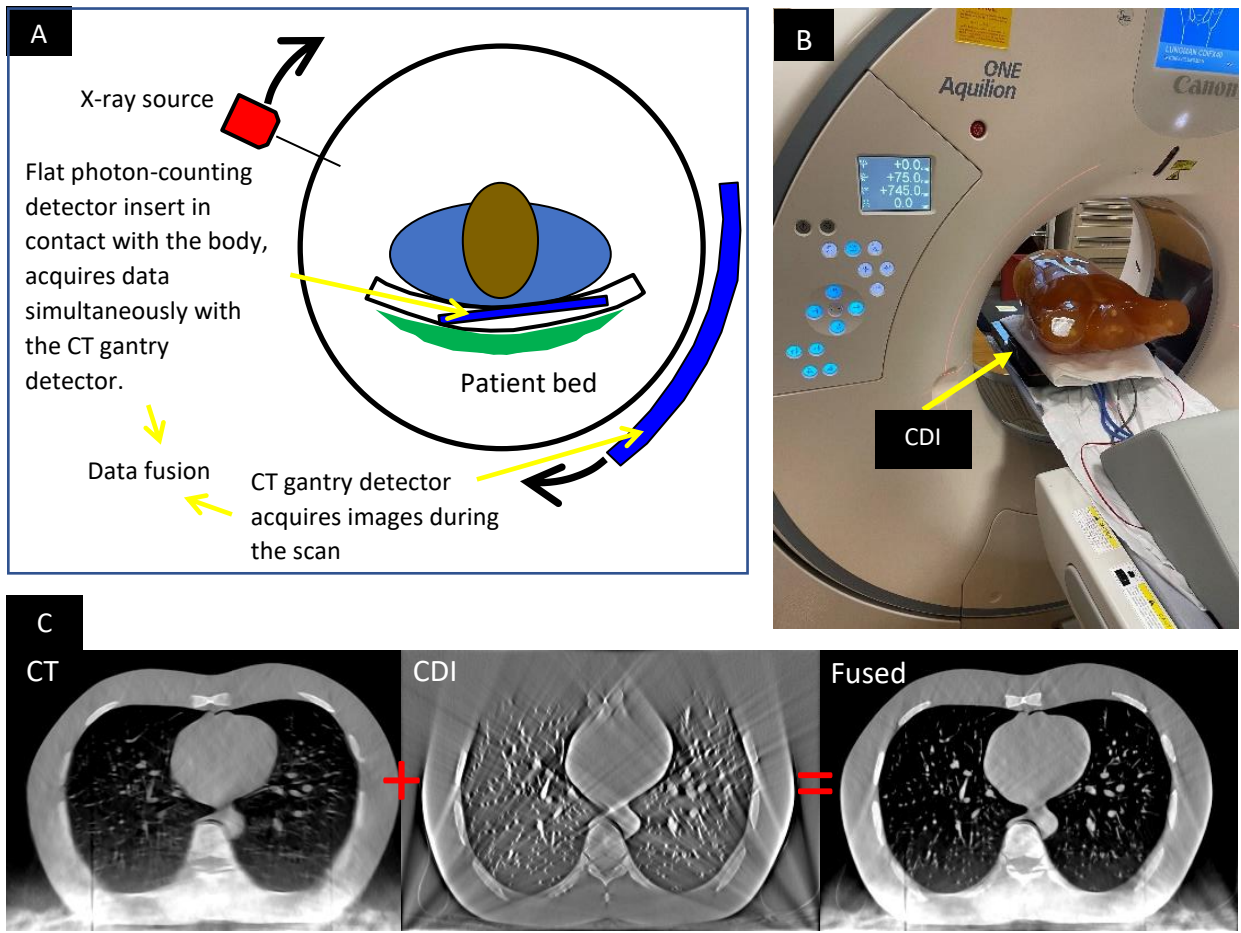

**Fig. 1.** (**A**) Schematic of the hybrid imaging system (**B**) Experimental hybrid system in a commercial CT scanner with the LUNGMAN chest phantom placed on the contact detector insert. (**C**) The images of the hybrid system are a weighted sum of the images from the CDI and the CT scanner. The CDI and CT images each contains missing information and artifacts from truncated or blocked projection angles. They are complementary to each other, such that the fused image has less artifacts and more complete depiction of the anatomical features.

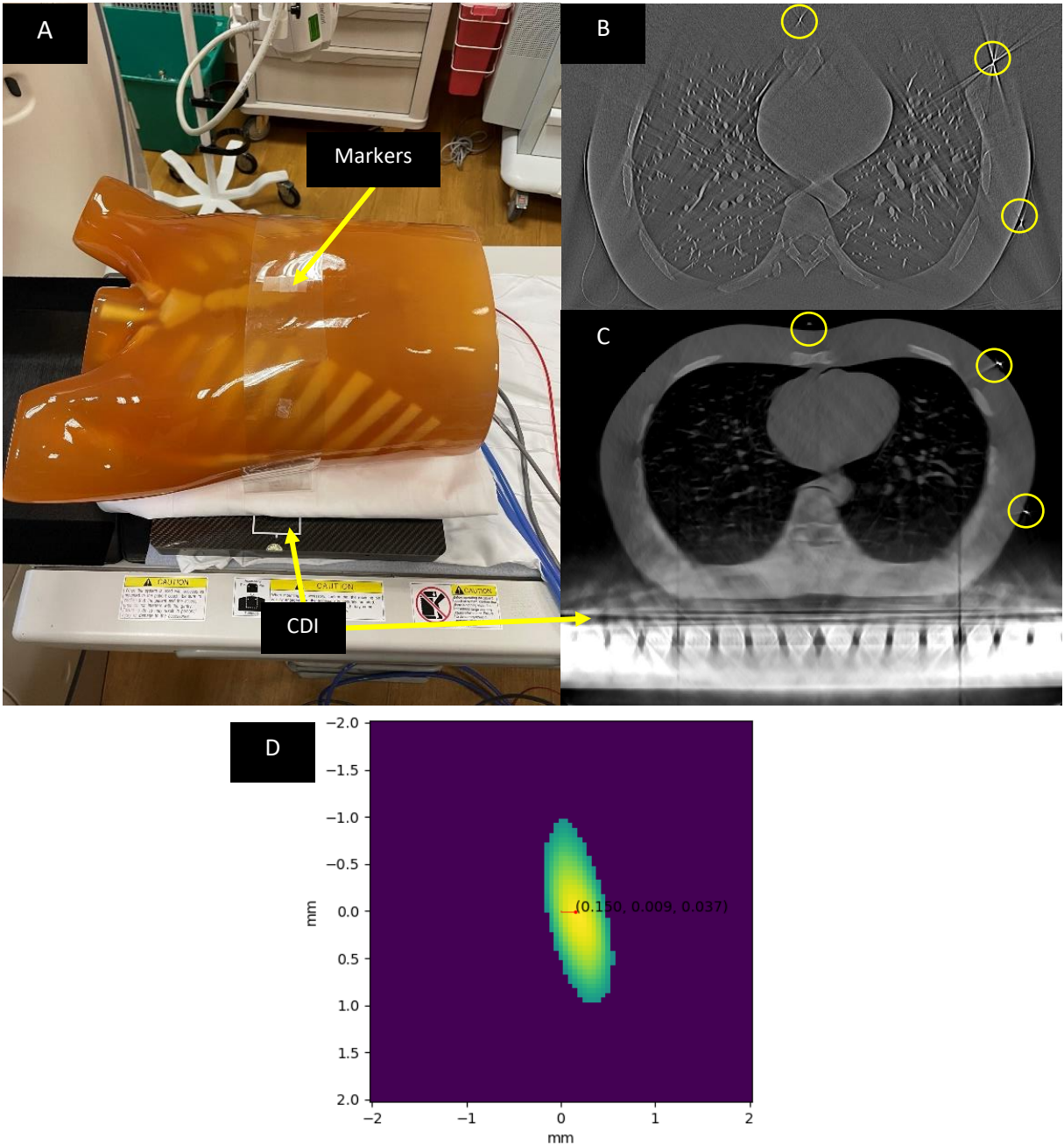

**Fig. 2.** (**A**) The LUNGMAN phantom laying on the CDI with fiducial markers placed on the outer surface of the phantom. (**B**) Slice of the CDI's reconstructed volume in the CT coordinate system. (**C**) The same slice in the CT images. Regions circled in yellow highlight visible fiducial markers. (**D**) Intensity profile of a marker bead in a CT image, after thresholding at half maximum intensity. The image was reconstructed at 50 µm pixel size. Center-of-mass of the profile is calculated as the location of a fiducial marker.

# Results

The calibration algorithm completed minimization of the cost function in 120 iterative steps. Figure 3 summarizes the quantitative results of the iterative procedure. The root-mean-squared (RMS) of the misalignment of the fiducial marker positions between the CDI and CT images was reduced by 68% with the procedure, from 0.563 mm to 0.182 mm. As a result, the RMS of the misalignment of small blood vessels between the two sets of images was reduced by 65%, from 0.375 mm to 0.131 mm. An example of improved alignment of small vessels is shown in Fig. 3.

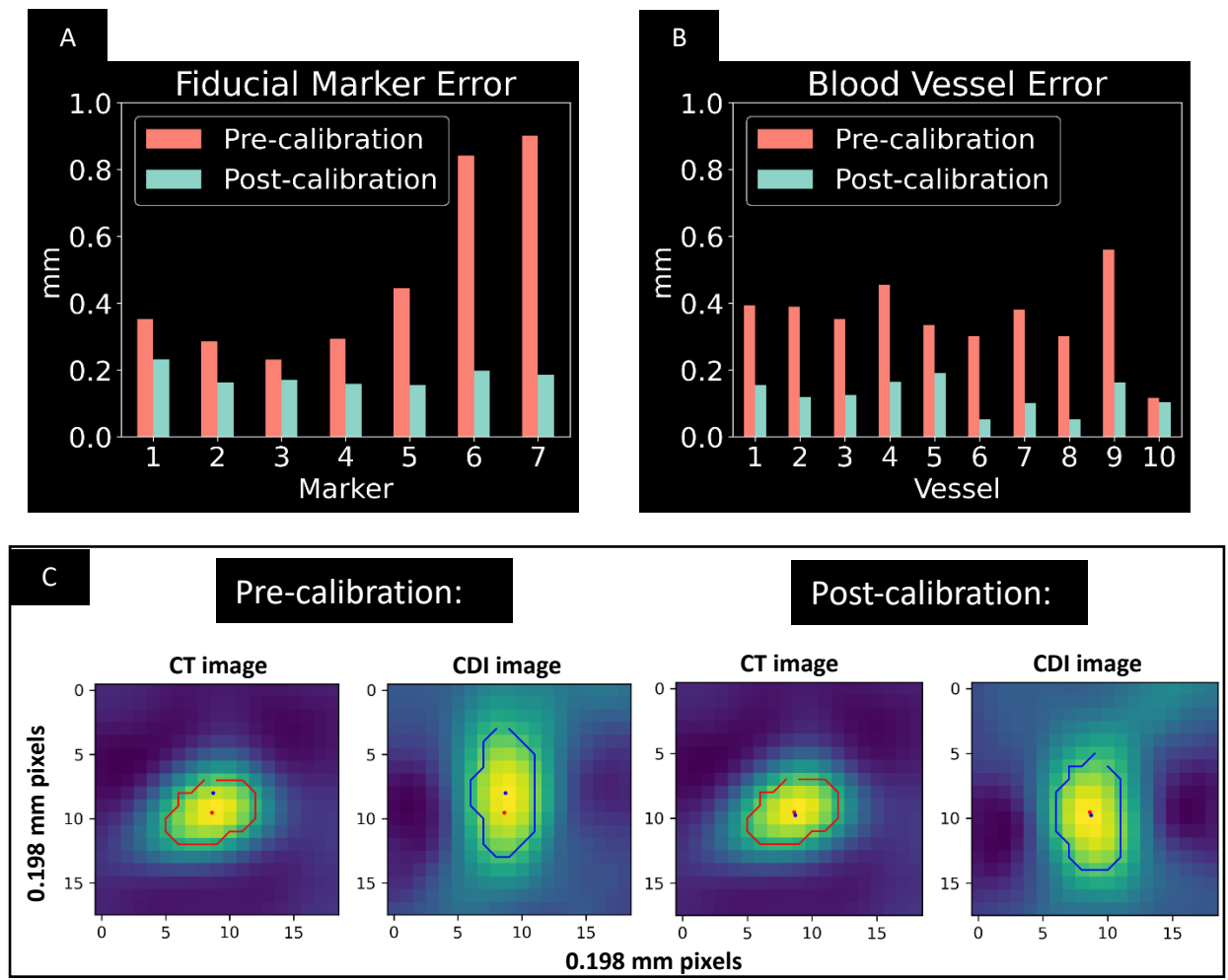

**Fig. 3.** (**A**) Pre- and post-calibration alignment errors of the marker beads between the CDI images and the CT images. (**B**) Pre- and post-calibration alignment errors of small blood vessels ( < 2mm diameter) between the two detectors. The vessels were distributed evenly in the reconstructed volume. (**C**) An example of a blood vessel position in the CT images (red contour) and CDI images (blue contour). The red and blue dots are the center-of-mass positions of the vessel in the CDI and CT images, respectively.

The overall effect of the iterative calibration procedure on the reconstructed images of the hybrid system is illustrated in magnified regions of the right and left lungs (Fig. 4). Fusion of the two sources of data with pre-calibrated parameters resulted in substantial streaking artifacts (Fig. 4D) and misalignment of features such as vessel segments from the two sources (Fig. 4B). After the iterative procedure, the streaking artifacts were reduced compared to pre calibration (Fig. 4E), and the vessel segments were aligned (Fig. 4C).

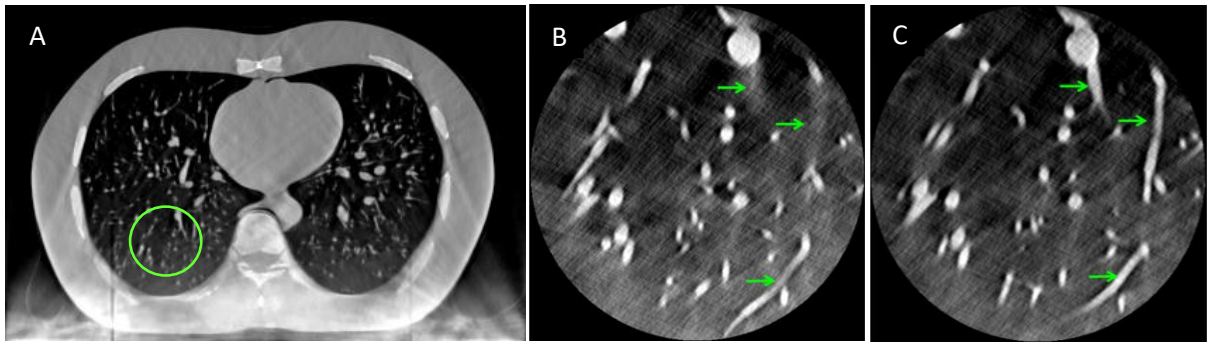

**Fig. 4. (A)** 500 mm FOV image of the chest phantom after data fusion. (**B**) and (**C**): pre- and post-calibration 50 mm FOV (0.098 mm pixel) CT-CDI fused images of a region highlighted by the green circle in (A). Arrows point to the difference in the visualization of vessel structures before and after the iterative calibration.

## Discussion

The hybrid CT system presented a unique geometric calibration problem and required on-line calibration for each scan. The method presented here is a hybridization of phantom-based calibration and image-based iterative optimization. As verified by fiducial marker and blood vessel alignment, the method was able to determine the geometric parameters of the system on a per-scan basis. The process led to a reduction in misalignment artifacts and enabled accurate fusion of the images from the CDI and the CT scanner. The resulting image from this fusion provided a more complete representation of the object, as the two datasets each contain complementary information.

Limitations of this calibration method arise from the statistical error in locating fiducial marker positions in the both the CT and CDI reconstructed volumes. The locations of the markers depended on the threshold value applied to the image before calculating the center of mass, the resolution of the reconstruction in all three dimensions, and the presence of image artifacts. The calculated marker positions fluctuated by about 0.1 mm between different threshold values. This level of uncertainty and the errors associated with tomosynthesis artifacts can explain the residual misalignment between the CDI and CT datasets after the iterative process (Fig. 3). Choosing tungsten carbide as the fiducial marker material helped to retain the majority of the marker intensity profile after thresholding the image due to the markers' high relative brightness.  However, choosing the threshold value is somewhat subjective and can lead to slightly different results; we found that using a consistent half-of-maximum intensity value was adequate to achieve desired calibration results in this initial study.

Further improvements of this calibration method will aim to shorten the total computation time required. As an online method, geometric calibration will need to be performed for each patient scan, and it would be ideal to reduce the total computation time down to seconds. This goal requires further code optimization in addition to larger parallel processing capability.  To improve alignment results, image artifacts in the CDI's reconstruction need to be reduced with refined reconstruction algorithms. Suppressing these artifacts may improve the intensity profile of the fiducial markers via reduced streaking, and in turn improve the precision of the calibration.

References

[1]     T. C. Larsen, E. E. Bennett, D. Mazilu, M. Y. Chen, and H. Wen, "Regional Ultrahigh-Resolution Rescan in a Clinical Whole-Body CT Scanner Using a Contact Detector Insert," *Tomography*, vol. 5, no. 2, Art. no. 2, Jun. 2019, doi: 10.18383/j.tom.2019.00002.

[2]     H. Wen *et al.*, "First Clinical Trial of a Photon-counting Surface Detector Insert in a Whole-body CT Scanner to Provide Resolution in Chest Scans Beyond the Scanner's Hardware Limits," in *RSNA Annual Meeting*, 2020, p. SSCH07.

[3]     J. T. Dobbins and D. J. Godfrey, "Digital x-ray tomosynthesis: current state of the art and clinical potential," *Phys. Med. Biol.*, vol. 48, no. 19, pp. R65–R106, Sep. 2003, doi: 10.1088/0031-9155/48/19/R01.

[4]     S. Vedantham, A. Karellas, G. R. Vijayaraghavan, and D. B. Kopans, "Digital Breast Tomosynthesis: State of the Art," *Radiology*, vol. 277, no. 3, pp. 663–684, Dec. 2015, doi: 10.1148/radiol.2015141303.

[5]     A. Chong, S. P. Weinstein, E. S. McDonald, and E. F. Conant, "Digital Breast Tomosynthesis: Concepts and Clinical Practice," *Radiology*, vol. 292, no. 1, pp. 1–14, Jul. 2019, doi: 10.1148/radiol.2019180760.

[6]     G. T. Gullberg, B. M. W. Tsui, C. R. Crawford, J. G. Ballard, and J. T. Hagius, "Estimation of geometrical parameters and collimator evaluation for cone beam tomography," *Med. Phys.*, vol. 17, no. 2, pp. 264–272, 1990, doi: 10.1118/1.596505.

[7] S. G. Azevedo, D. J. Schneberk, J. P. Fitch, and H. E. Martz, "Calculation of the rotational centers in computed tomography sinograms," *IEEE Trans. Nucl. Sci.*, vol. 37, no. 4, pp. 1525–1540, Aug. 1990, doi: 10.1109/23.55866.

[8] F. Noo, R. Clackdoyle, C. Mennessier, T. A. White, and T. J. Roney, "Analytic method based on identification of ellipse parameters for scanner calibration in cone-beam tomography," *Phys. Med. Biol.*, vol. 45, no. 11, pp. 3489–3508, 2000.

[9] G. M. Stevens, R. Saunders, and N. J. Pelc, "Alignment of a volumetric tomography system," *Med. Phys.*, vol. 28, no. 7, pp. 1472–1481, 2001, doi: 10.1118/1.1382609.

[10] D. Beque, J. Nuyts, G. Bormans, P. Suetens, and P. Dupont, "Characterization of pinhole SPECT acquisition geometry," *IEEE Trans. Med. Imaging*, vol. 22, no. 5, pp. 599–612, May 2003, doi: 10.1109/TMI.2003.812258.

[11] L. von Smekal, M. Kachelriess, E. Stepina, and W. A. Kalender, "Geometric misalignment and calibration in cone-beam tomography," *Med. Phys.*, vol. 31, no. 12, pp. 3242–3266, Dec. 2004, doi: 10.1118/1.1803792.

[12] Y. Cho, D. J. Moseley, J. H. Siewerdsen, and D. A. Jaffray, "Accurate technique for complete geometric calibration of cone-beam computed tomography systems," *Med. Phys.*, vol. 32, no. 4, pp. 968–983, 2005, doi: 10.1118/1.1869652.

[13] K. Yang, A. L. C. Kwan, D. F. Miller, and J. M. Boone, "A geometric calibration method for cone beam CT systems," *Med. Phys.*, vol. 33, no. 6Part1, pp. 1695–1706, 2006, doi: 10.1118/1.2198187.

[14] X. Wang, J. G. Mainprize, M. P. Kempston, G. E. Mawdsley, and M. J. Yaffe, "Digital breast tomosynthesis geometry calibration," in *Medical Imaging 2007: Physics of Medical Imaging*, Mar. 2007, vol. 6510, pp. 1118–1128. doi: 10.1117/12.698714.

[15] S. Hoppe, F. Noo, F. Dennerlein, G. Lauritsch, and J. Hornegger, "Geometric calibration of the circle-plus-arc trajectory," *Phys. Med. Biol.*, vol. 52, no. 23, pp. 6943–6960, Nov. 2007, doi: 10.1088/0031-9155/52/23/012.

[16] D. Panetta, N. Belcari, A. D. Guerra, and S. Moehrs, "An optimization-based method for geometrical calibration in cone-beam CT without dedicated phantoms," *Phys. Med. Biol.*, vol. 53, no. 14, pp. 3841–3861, Jun. 2008, doi: 10.1088/0031-9155/53/14/009.

[17] Y. Kyriakou, R. M. Lapp, L. Hillebrand, D. Ertel, and W. A. Kalender, "Simultaneous misalignment correction for approximate circular cone-beam computed tomography," *Phys. Med. Biol.*, vol. 53, no. 22, pp. 6267–6289, Nov. 2008, doi: 10.1088/0031-9155/53/22/001.

[18] V. Patel, R. N. Chityala, K. R. Hoffmann, C. N. Ionita, D. R. Bednarek, and S. Rudin, "Self-calibration of a cone-beam micro-CT system," *Med. Phys.*, vol. 36, no. 1, pp. 48–58, Jan. 2009, doi: 10.1118/1.3026615.

[19] X. Li, D. Zhang, and B. Liu, "A generic geometric calibration method for tomographic imaging systems with flat-panel detectors—A detailed implementation guide," *Med. Phys.*, vol. 37, no. 7Part1, pp. 3844–3854, 2010, doi: 10.1118/1.3431996.

[20] A. Kingston, A. Sakellariou, T. Varslot, G. Myers, and A. Sheppard, "Reliable automatic alignment of tomographic projection data by passive auto-focus," *Med. Phys.*, vol. 38, no. 9, pp. 4934–4945, 2011, doi: 10.1118/1.3609096.

[21] X. Li, D. Zhang, and B. Liu, "Sensitivity analysis of a geometric calibration method using projection matrices for digital tomosynthesis systems," *Med. Phys.*, vol. 38, no. 1, pp. 202–209, 2011, doi: 10.1118/1.3524221.

[22] D. Wu, L. Li, L. Zhang, Y. Xing, Z. Chen, and Y. Xiao, "Geometric calibration of cone-beam CT with a flat-panel detector," in *2011 IEEE Nuclear Science Symposium Conference Record*, Oct. 2011, pp. 2952–2955. doi: 10.1109/NSSMIC.2011.6152527.

[23] S. Sawall, M. Knaup, and M. Kachelrieß, "A robust geometry estimation method for spiral, sequential and circular cone-beam micro-CT," *Med. Phys.*, vol. 39, no. 9, pp. 5384–5392, 2012, doi: 10.1118/1.4739506.

[24] D. Gross, U. Heil, R. Schulze, E. Schoemer, and U. Schwanecke, "Auto calibration of a cone-beam-CT," *Med. Phys.*, vol. 39, no. 10, pp. 5959–5970, 2012, doi: 10.1118/1.4739247.

[25] J. Wicklein, H. Kunze, W. A. Kalender, and Y. Kyriakou, "Image features for misalignment correction in medical flat-detector CT," *Med. Phys.*, vol. 39, no. 8, pp. 4918–4931, 2012, doi: 10.1118/1.4736532.

[26] H. Miao, X. Wu, H. Zhao, and H. Liu, "A phantom-based calibration method for digital x-ray tomosynthesis," *J. X-Ray Sci. Technol.*, vol. 20, no. 1, pp. 17–29, Jan. 2012, doi: 10.3233/XST-2012-0316.

[27] A. Ladikos and W. Wein, "Geometric calibration using bundle adjustment for cone-beam computed tomography devices," in *Medical Imaging 2012: Physics of Medical Imaging*, Mar. 2012, vol. 8313, pp. 814–819. doi: 10.1117/12.906238.

[28] Y. Meng, H. Gong, and X. Yang, "Online Geometric Calibration of Cone-Beam Computed Tomography for Arbitrary Imaging Objects," *IEEE Trans. Med. Imaging*, vol. 32, no. 2, pp. 278–288, Feb. 2013, doi: 10.1109/TMI.2012.2224360.

[29] I. Ben Tekaya, V. Kaftandjian, F. Buyens, S. Sevestre, and S. Legoupil, "Registration-Based Geometric Calibration of Industrial X-ray Tomography System," *IEEE Trans. Nucl. Sci.*, vol. 60, no. 5, pp. 3937–3944, Oct. 2013, doi: 10.1109/TNS.2013.2279675.

[30] M. Xu, C. Zhang, X. Liu, and D. Li, "Direct determination of cone-beam geometric parameters using the helical phantom," *Phys. Med. Biol.*, vol. 59, no. 19, pp. 5667–5690, Sep. 2014, doi: 10.1088/0031-9155/59/19/5667.

[31] A. Zechner *et al.*, "Development and first use of a novel cylindrical ball bearing phantom for 9-DOF geometric calibrations of flat panel imaging devices used in image-guided ion beam therapy," *Phys. Med. Biol.*, vol. 61, no. 22, pp. N592–N605, Oct. 2016, doi: 10.1088/0031-9155/61/22/N592.

[32] K. Zhou *et al.*, "A New Method for Cone-Beam Computed Tomography Geometric Parameters Estimation," *J. Comput. Assist. Tomogr.*, vol. 40, no. 4, pp. 639–648, Aug. 2016, doi: 10.1097/RCT.0000000000000393.

[33]     M. W. Jacobson *et al.*, "A line fiducial method for geometric calibration of cone-beam CT systems with diverse scan trajectories," *Phys. Med. Biol.*, vol. 63, no. 2, p. 025030, Jan. 2018, doi: 10.1088/1361-6560/aa9910.

[34]     C. Jiang, N. Zhang, J. Gao, and Z. Hu, "Geometric calibration of a stationary digital breast tomosynthesis system based on distributed carbon nanotube X-ray source arrays," *PLOS ONE*, vol. 12, no. 11, p. e0188367, Nov. 2017, doi: 10.1371/journal.pone.0188367.

[35]     C. J. Choi, T. L. Vent, R. J. Acciavatti, and A. D. A. Maidment, "Geometric calibration for a next-generation digital breast tomosynthesis system using virtual line segments," in *Medical Imaging 2018: Physics of Medical Imaging*, Mar. 2018, vol. 10573, pp. 89–98. doi: 10.1117/12.2294634.

[36]     G. Li *et al.*, "A novel calibration method incorporating nonlinear optimization and ball-bearing markers for cone-beam CT with a parameterized trajectory," *Med. Phys.*, vol. 46, no. 1, pp. 152–164, 2019, doi: 10.1002/mp.13278.

[37]     C.-H. Chang, Y.-C. Ni, S.-Y. Huang, H.-H. Hsieh, S.-P. Tseng, and F.-P. Tseng, "A geometric calibration method for the digital chest tomosynthesis with dual-axis scanning geometry," *PLOS ONE*, vol. 14, no. 4, p. e0216054, Apr. 2019, doi: 10.1371/journal.pone.0216054.

[38]     V. Nguyen, J. G. Sanctorum, S. Van Wassenbergh, J. J. J. Dirckx, J. Sijbers, and J. De Beenhouwer, "Geometry Calibration of a Modular Stereo Cone-Beam X-ray CT System," *J. Imaging*, vol. 7, no. 3, Art. no. 3, Mar. 2021, doi: 10.3390/jimaging7030054.

[39]     J. Graetz, "Auto-calibration of cone beam geometries from arbitrary rotating markers using a vector geometry formulation of projection matrices," *Phys. Med. Biol.*, vol. 66, no. 7, p. 075013, Apr. 2021, doi: 10.1088/1361-6560/abe75f.

[40]     S. Moon, S. Choi, H. Jang, M. Shin, Y. Roh, and J. Baek, "Geometry calibration and image reconstruction for carbon-nanotube-based multisource and multidetector CT," *Phys. Med. Biol.*, vol. 66, no. 16, p. 165005, Aug. 2021, doi: 10.1088/1361-6560/ac16c1.

[41]     X. Duan *et al.*, "Knowledge-based self-calibration method of calibration phantom by and for accurate robot-based CT imaging systems," *Knowl.-Based Syst.*, vol. 229, p. 107343, Oct. 2021, doi: 10.1016/j.knosys.2021.107343.

[42]     M. J. D. Powell, "An efficient method for finding the minimum of a function of several variables without calculating derivatives," *Comput. J.*, vol. 7, no. 2, pp. 155–162, Jan. 1964, doi: 10.1093/comjnl/7.2.155.